\documentclass[Journal]{IEEEtran}

\usepackage{graphicx} 
\usepackage{subfigure}
\usepackage{epstopdf}
\usepackage[square,sort&compress,comma,numbers]{natbib}
\usepackage{color}
\usepackage{amsfonts}
\usepackage{latexsym}
\usepackage{amssymb}
\usepackage{amsmath}
\usepackage{multirow}
\usepackage{upgreek}
\usepackage{algorithm}
\usepackage{algpseudocode}
\usepackage{mathrsfs}
\usepackage{threeparttable}
\usepackage{bm}
\renewcommand{\algorithmicrequire}{\textbf{Input:}}  
 
\renewcommand{\algorithmicrepeat}{\textbf{Find the artifact related dimensions in cepstral domain:}}

\usepackage[table,xcdraw]{xcolor}

\usepackage{tikz,xcolor,hyperref}
\definecolor{lime}{HTML}{A6CE39}
\DeclareRobustCommand{\orcidicon}{%
    \begin{tikzpicture}
    \draw[lime, fill=lime] (0,0) 
    circle [radius=0.16] 
    node[white] {{\fontfamily{qag}\selectfont \tiny ID}};    \draw[white, fill=white] (-0.0625,0.095) 
    circle [radius=0.007];    \end{tikzpicture}
    \hspace{-2mm}}
\foreach \x in {A, ..., Z}{%
    \expandafter\xdef\csname orcid\x\endcsname{\noexpand\href{https://orcid.org/\csname orcidauthor\x\endcsname}{\noexpand\orcidicon}}
    }
\hypersetup{hidelinks,
	colorlinks=true,
	allcolors=black,
	pdfstartview=Fit,
	breaklinks=true}

\IEEEoverridecommandlockouts
\begin{document}
\title{Cepstral Analysis Based Artifact Detection, Recognition and Removal for Prefrontal EEG}

\author{\IEEEauthorblockN
        {Siqi Han\IEEEauthorrefmark{1},
        Chao Zhang\IEEEauthorrefmark{1}\orcidA{},
        Jiaxin Lei\orcidC{},
        Qingquan Han,
        Yuhui Du,
        Anhe Wang,
        Shuo Bai,\\
        Milin Zhang\orcidB{},~\IEEEmembership{Senior Member,~IEEE}
        }

\thanks{\IEEEauthorrefmark{1}Siqi Han and Chao Zhang contributed equally to this work.}
\thanks{Siqi Han is with the School of Modern Post (School of Automation), Beijing University of Posts and Telecommunications, Beijing, China, 100876.}
\thanks{Chao Zhang, Jiaxin Lei and Yuhui Du are with the Department of Electronic Engineering, Tsinghua University, Beijing, China, 100084.}
\thanks{Qingquan Han, Anhe Wang and Shuo Bai are with State Key Laboratory of Biochemical Engineering, Institute of Process Engineering, Chinese Academy of Sciences, Beijing, China, 100084.}
\thanks{Milin Zhang is with the Department of Electronic Engineering, Beijing National Research Center for Information Science and Technology, Institute for Precision Medicine, Tsinghua University, Beijing, China, 100084. Corresponding author e-mail: zhangmilin@tsinghua.edu.cn}
\thanks{This work is supported in part by the National Natural Science Foundation of China (Nos. NSFC62227801), in part by the Natural Science Foundation of China through grant 92164202, in part by the Beijing Innovation Center for Future Chip.}
}

\maketitle
\begin{abstract}
This paper proposes to use cepstrum for artifact detection, recognition and removal in prefrontal EEG.
This work focuses on the artifact caused by eye movement. A database containing artifact-free EEG and eye movement contaminated EEG from different subjects is established.
A cepstral analysis-based feature extraction with support vector machine (SVM) based classifier is designed to identify the artifacts from the target EEG signals.
The proposed method achieves an accuracy of 99.62\% on the artifact detection task and a 82.79\% accuracy on the 6-category eye movement classification task. 
A statistical value-based artifact removal method is proposed and evaluated on a public EEG database, where an accuracy improvement of 3.46\% is obtained on the 3-category emotion classification task.
In order to make a confident decision of each 5s EEG segment, the algorithm requires only 0.66M multiplication operations.
Compared to the state-of-the-art approaches in artifact detection and removal, the proposed method features higher detection accuracy and lower computational cost, which makes it a more suitable solution to be integrated into a real-time and artifact robust Brain-Machine Interface (BMI).
\end{abstract}

\begin{IEEEkeywords}
Brain-machine interface (BMI), Artifact detection and removal, Eye movement, Cepstral analysis, Support vector machine (SVM)
\end{IEEEkeywords}

\IEEEpeerreviewmaketitle

\section{Introduction}
Electroencephalogram (EEG) is widely used in various Brain-Machine Interface (BMI) applications and clinical practice for nervous system disorder diagnosis \cite{tang-nature,rssi-seizure,mcmd-sleep}. The amplitude of the EEG signal is in the range of several micro volts. It can be easily interfered by various noises, such as powerline interference, the instability of the bio-ground in the wearable acquisition device, unexpected movement and/or vibration, cable sway, electromyography (EMG) and/or electrooculography (EOG) artifacts \cite{jiang2019removal-review}.
The EMG/EOG artifacts usually feature millivolts in amplitude, which make it a disaster in prefrontal EEG acquisition and analysis.
If an artifact is not properly removed from the acquired EEG signals, the output decision of the EEG-driven system may be misleading.

Various artifact detection and removal algorithms have been proposed in literature \cite{zou2019removing,mahajan-ica-JBHI2014,daly2014force,lee2018removal,cimmino-pca-2021,asr-tcasii,chen-cca-2018}. 
Based on the assumption that EEG and the artifacts are uncorrelated, blind source separation (BSS) methods are widely utilized, such as independent component analysis (ICA) \cite{mahajan-ica-JBHI2014,daly2014force}, independent vector analysis (IVA) \cite{lee2018removal}, principal component analysis (PCA) \cite{cimmino-pca-2021,asr-tcasii} and canonical correlation analysis (CCA) \cite{chen-cca-2018}. 
In addition, adaptive filter \cite{2019-filter}, discrete wavelet transform (DWT) \cite{khatun2016-wavelet,2020-JBHI-wavelet}, Fourier-Bessel series expansion based empirical wavelet transform (FBSE-EWT) \cite{gajbhiye2019elimination} and deep learning \cite{khatwani-biocas18,manjunath2020low,deep-learning2020} have also been introduced to perform artifact detection and/or removal. 
The existing solutions can achieve competitive artifact detection sensitivity and offer a rich arsenal for artifact removal, but suffer from the high complexity and/or high latency which makes them unsuitable for portable devices. On the other hand, few works have studied the identification between the different origins of the artifacts, which consists of valuable information of specific action.

Inspired from the success of cepstral features \cite{childers1977cepstrum} such as Mel-Frequency Cepstral Coefficients (MFCCs) in the field of speech signal processing \cite{juvela2018-ICASSP18}, this paper proposes to apply cepstral feature extraction into artifact detection, the artifact resources recognition and the artifact removal.
Such as the frequency spectrum describes the rhythm of the time domain series, the cepstral coefficients describe the rhythm of the frequency domain series.
The artifacts are detected from the recorded EEG signals according to the cepstral features based analysis. The artifact resources can be further classified into five categories according to different eye movements. In addition, an artifact removal algorithm in cepstral domain is proposed and evaluated on a public EEG database.

The rest of the paper is organized as follows. 
Section II introduces the EEG recording setup as well as the proposed cepstral analysis method. 
Section III illustrates the experimental results in artifact detection, recognition and removal, while Section IV concludes the entire work.

\section{Data set and Cepstral analysis}

\subsection{Device}
\begin{figure}[htb]
  \centering
  \includegraphics[width=0.48\textwidth]{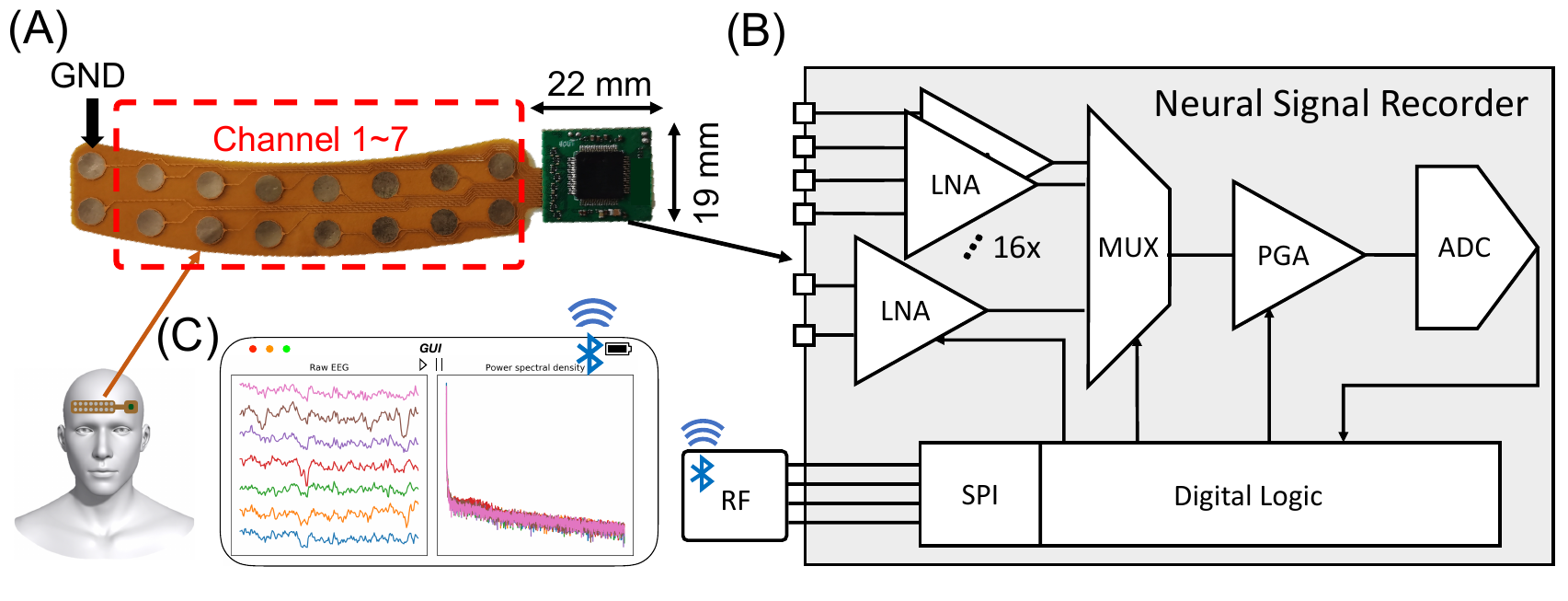}\\
  \caption{The prefrontal EEG recording system. (A) The FPC based multi-channel electrode and the recorder. (B) The diagram of the recorder. (C) The GUI on an iPad.}
  \label{fig:system}
\end{figure}

Fig. \ref{fig:system} shows the architecture of the proposed prefrontal EEG acquisition system. It consists of 1) a flexible print circuit (FPC) based multiple channel EEG electrode, 2) a wireless recorder \cite{luo2019low} with a size of 19 mm $\times$ 22 mm $\times$ 2 mm and a weight of 2.6 g, and 3) a graphical user interface (GUI).
The FPC is customized to adapt the size and shape of the forehead of adults. Sixteen hydrogel based electrodes are placed on the FPC. There are 7 pairs of differential EEG channels and one pair of ground (GND) channel. 
The recorder amplifies and quantifies the acquired EEG signals.
The digitized EEG signals are sent to the GUI wirelessly through BlueTooth. The waveform of the acquired raw data as well as the calculated spectrums are plotted on the GUI in real-time.

\subsection{Data Acquisition}
Five subjects, with 4 males and 1 female, with an average age of 22.8$\pm$1.6 years old were recruited for the EEG recording.
The data acquisition was conducted in a closed room.
In each set of the experiments, the subjects were asked to perform five actions, denoted as blinking hard, looking upward, looking downward, looking leftward and looking rightward.
Each action lasts for 20 seconds. There is a 10 seconds break between every two actions. The natural blink is allowed in all recording process to align with the real-lift scene and avoid making subjects feel uncomfortable.
The experiment was divided into five trials. The sequence of the actions in each group was different to ensure that each action appeared once in each position. 
In each trial, the subject was asked to close eyes for 20 seconds. Then the action sequence was executed twice.
Each trial lasted for 5 minutes 20 seconds. A total 26 minutes 40 seconds EEG was recorded from each subject.

The acquired EEG series were separated into 5-second intervals. Each interval was regarded as a single sample. Focusing on the heavy artifacts which would destroy the quality of prefrontal EEG, the EEG data in the relax state was considered as samples without artifacts. In contrast, the EEG signals of the 5 designed actions were considered as five different artifact origins.

\subsection{Cepstral Analysis}

\begin{figure}[htb]
  \centering
  \includegraphics[width=0.48\textwidth]{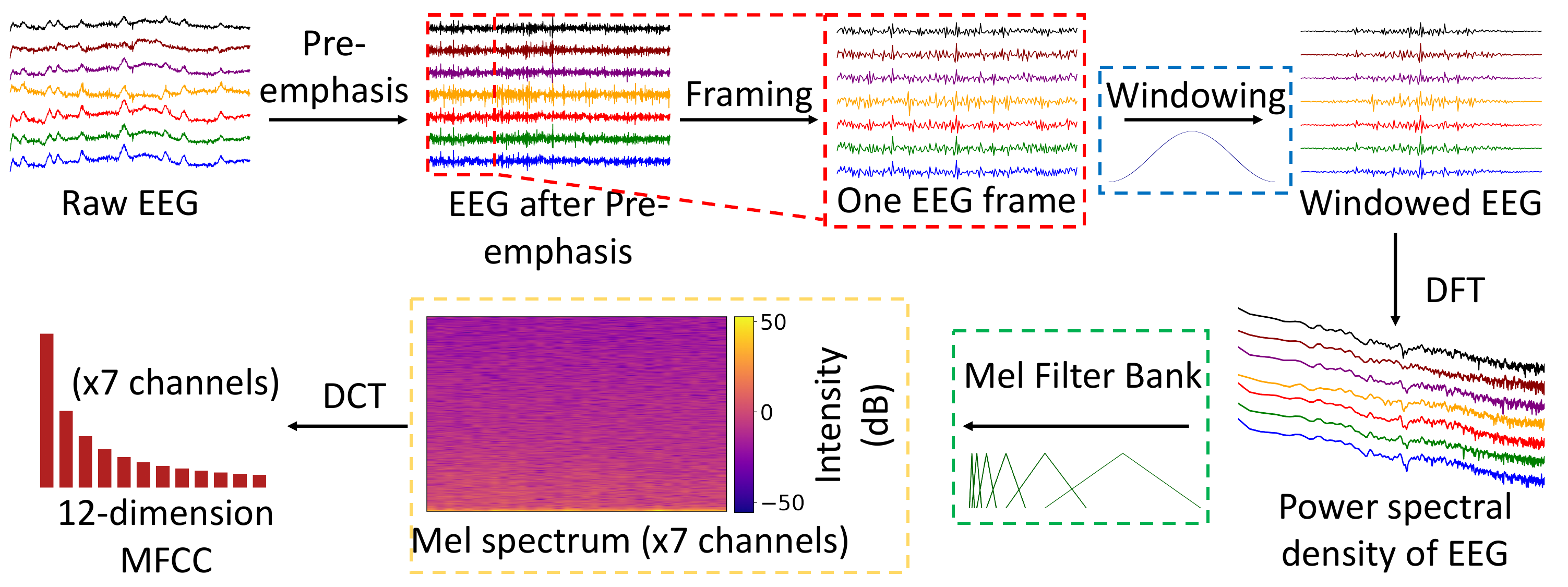}\\
  \caption{The process flow of MFCCs.}
  \label{fig:process-MFCC}
\end{figure}

Cepstral analysis is a nonlinear signal processing technique widely used in areas such as speech processing.
In this work, Mel-Frequency Cepstral Coefficients (MFCCs) \cite{mfcc1980} as shown in Fig. \ref{fig:process-MFCC} were exploited for the deconvolution of the target EEG and the unexpected artifacts.

The sources of the target EEG and the unexpected artifacts are different from the position where the contaminated EEG is recorded, which means artifacts could influence the transition paths from the original EEG to the electrodes.
The recorded EEG, $x(n)$, can be modeled as a convolution of the target EEG, $s(n)$, and the artifact, $h(n)$, as 
\begin{equation}
\label{eq:cesp-1}
\begin{aligned}
x(n)=s(n)*h(n)
\end{aligned}
\end{equation}

\noindent {The first step of the calculation of MFCCs is pre-emphasis.
A high pass filter is applied to the EEG signal to increase the signal energy at higher frequencies as} 
\begin{equation}
\label{eq:emphasis}
\begin{aligned}
c(n)=b(n)-\beta b(n-1)
\end{aligned}
\end{equation}
where $b(n)$ is the input signal. $c(n)$ is the output signal. A typical value of 0.95 is applied for the constant $\beta$.
For the frequency domain representations $X(k)$ of $x(n)$, $S(k)$ of $s(n)$, $H(k)$ of $h(n)$, the pre-emphasis can be denoted as
\begin{equation}
\label{eq:cesp-emphasis-fre1}
\begin{aligned}
X(k)F(k)=S(k)H(k)F(k), 0\leq k\leq N-1
\end{aligned}
\end{equation}
where $F(k)$ is the frequency domain representation of the high pass filter. The frequency domain result $X_1(k)$ of the filtering is denoted as 
\begin{equation}
\label{eq:cesp-emphasis-fre2}
\begin{aligned}
X_1(k)=S(k)F_1(k)H(k)F_2(k)
\end{aligned}
\end{equation}
where $F_1(k)$ and $F_2(k)$ are two high pass filters.
\begin{equation}
\label{eq:cesp-emphasis-fre2-5}
\begin{aligned}
{\exists} \ F_1(k), F_2(k) \qquad  s.t. \ F(k)=F_1(k)F_2(k)
\end{aligned}
\end{equation}

\noindent{Eq.(4) is equivalently denoted as}
\begin{equation}
\label{eq:cesp-emphasis-fre3}
\begin{aligned}
X_1(k)=S_1(k)H_1(k)
\end{aligned}
\end{equation}
where $S_1(k)=S(k)F_1(k)$ and $H_1(k)=H(k)F_2(k)$.

Eq.(6) can be converted to Eq.(7) by iFFT. After this step, the pre-emphasised signals of Eq.(1) can be denoted as
\begin{equation}
\label{eq:cesp-emphasis-fre4}
\begin{aligned}
x_1(n)=s_1(n)*h_1(n)
\end{aligned}
\end{equation}
where $x_1(n)$, $s_1(n)$, $h_1(n)$ are the time domain representations of $X_1(k)$, $S_1(k)$, $H_1(k)$ respectively.

The second step is framing,
which will segment EEG signals into small frames with a specific length $N$. In this work, $N$ is set to be 2048. A single frame is donated as:
\begin{equation}
\label{eq:cesp-3}
\begin{aligned}
x_2(n)=s_2(n)*h_2(n), n=0,1,...,N-1
\end{aligned}
\end{equation}
where $x_2(n)$, $s_2(n)$, $h_2(n)$ are the outputs of the framing operation of $x_1(n)$, $s_1(n)$, $h_1(n)$ respectively.

A single frame is used in the following derivation to obtain MFCCs. The MFCCs of a certain EEG segment are calculated by averaging the MFCCs of several frames in the segment.

The third step is windowing, in which a window is used to reduce the amplitude of the discontinuities at the boundaries of each finite sequence.
A Hamming window $w(n)$ is used in this work.
The output signal $x_3(n)$ of the third step is denoted as:
\begin{equation}
\label{eq:cesp-4}
\begin{aligned}
x_3(n)=w(n)s_2(n)*h_2(n)=s_3(n)*h_3(n), \\n=0,1,...,N-1
\end{aligned}
\end{equation}
where $s_3(n)$ and $h_3(n)$ are the windowed signals of $s_2(n)$ and $h_2(n)$, respectively.

\noindent {Discrete Fourier Transform (DFT) is used to convert each frame from time domain into frequency domain as }
\begin{equation}
\label{eq:dft}
\begin{aligned}
{X}(k)=\sum_{n=0}^{N-1}{x}(n)w(n)e^{-\frac{j2\pi kn}{N}},0\leq k\leq N-1
\end{aligned}
\end{equation}
The frequency domain representation is denoted as:
\begin{equation}
\label{eq:cesp-5}
\begin{aligned}
X_3(k)=S_3(k)H_3(k)
\end{aligned}
\end{equation}
where $X_3(k)$, $S_3(k)$, $H_3(k)$ are the frequency domain representations of $x_3(n)$, $s_3(n)$, $h_3(n)$ respectively.

The next step is Mel filter bank processing.
Mel Filter Bank usually is a set of 20-40 (40 in this work) triangular filters.
The magnitude frequency response of each Mel Filter equals to 1 at its centre frequency. It decreases linearly to 0 at the centre frequencies of the two adjacent filters. 
Then each Mel Filter is multiplied with the amplitude spectrum $|X_3(k)|$.
The Mel scale is calculated for given frequency f in Hz as
\begin{equation}
\label{eq:mel}
\begin{aligned}
F(Mel)=2595\times{log_{10}}(1+\frac{f}{700})
\end{aligned}
\end{equation}
After this step, the frequency spectrum is defined as
\begin{equation}
\label{eq:cesp-6}
\begin{aligned}
X_4(k)=S_4(k)H_4(k)
\end{aligned}
\end{equation}

\noindent{where $X_4(k)$, $S_4(k)$, $H_4(k)$ are the outputs of the Mel filter bank processing of $X_3(k)$, $S_3(k)$, $H_3(k)$ respectively.}
\noindent{The log Mel Spectrum can be calculated as}
\begin{equation}
\label{eq:cesp-log}
\begin{aligned}
\hat{X}(k)=log(X_4(k))&=log(S_4(k))+log(H_4(k))\\&=\hat{S}(k)+\hat{H}(k)
\end{aligned}
\end{equation}

\noindent{where $\hat{X}(k)$, $\hat{S}(k)$, $\hat{H}(k)$ is the logarithms of the Mel spectrums $X_4(k)$, $S_4(k)$, $H_4(k)$, respectively. The log Mel spectrum $\hat{X}(k)$ is converted into cepstral domain $\hat{x}(n)$ using Discrete Cosine Transform (DCT) as}
\begin{equation}
\label{eq:cesp-dct}
\begin{aligned}
\hat{x}(n)=\hat{s}(n)+\hat{h}(n)
\end{aligned}
\end{equation}
\noindent{where $\hat{s}(n)$ and $\hat{h}(n)$ are the DCT results of $\hat{S}(k)$ and $\hat{H}(k)$.}

The 2nd-13th dimensions of the DCT conversion result are denoted as MFCCs in this work. 
The deconvolution of $s(n)$ and $h(n)$ is accomplished through the spectral transform and the logarithm operation.
Typically, when the frequency distributions of $s(n)$ and $h(n)$ are different, the cepstral domain representations $\hat{s}(n)$ and $\hat{h}(n)$ hold few overlaps.

\section{Experiments}
\subsection{Artifact detection and recognition}
The MFCCs were firstly recruited to perform a Kruskal-Wallis test. The results showed there was a significant difference in each dimension of the MFCCs between the contaminated and the clean EEG samples ($p < 0.01$). Then the MFCCs were fed into a Radial Basis Function (RBF) kernel based SVM classifier for training. 
The classifier was tested by a 5-fold cross-validation (CV) protocol. 

In artifact detection experiment, the five actions are used as different artifact resources.
The samples acquired during the relaxing are used as no artifact samples. A binary classifier is trained to detect the existing of artifacts. 
An accuracy of 99.62$\%$ is measured in the 5-fold CV as shown in Table \ref{tb:metrics}, where the precision, recall and F1-score are also reported.
\begin{table}[!ht]
\caption{Performance Evaluation of the Proposed Method}
\label{tb:metrics}
\begin{center}
\scriptsize
\scalebox{1.0}{
\begin{tabular}{|c|c|c|}
\hline
& Artifact detection & Artifact recognition \\
\hline
Accuracy & 99.6\% $\pm$ 0.4\%  & 82.8\% $\pm$ 1.5\% \\
\hline
Precision & 99.6\% $\pm$ 0.4\% & 83.0\% $\pm$ 1.4\% \\
\hline
Recall & 99.6\% $\pm$ 0.4\% & 82.8\% $\pm$ 1.5\% \\
\hline
F1-score & 0.996 $\pm$ 0.004 & 0.829 $\pm$ 0.015\\
\hline
\end{tabular}}
\end{center}
\end{table}







In the artifact recognition experiment, samples of each action are used as input to train a 6-class classifier. 
Although subjects were asked to do the same eye movements, the same actions performed by different subjects are not exactly the same. For example, while performing the action of ``blinking hard", some subjects blinked twice at a time, and others blinked three times.
Due to the individual differences, compare with the binary classification, the performance of the classifier has declined, but still achieved an satisfactory 6-category classification accuracy of 82.79\% as shown in Table \ref{tb:metrics}. 

\subsection{Comparison between cepstrum and frequency spectrum}
\begin{figure}[htb]
  \centering
  \includegraphics[width=0.48\textwidth]{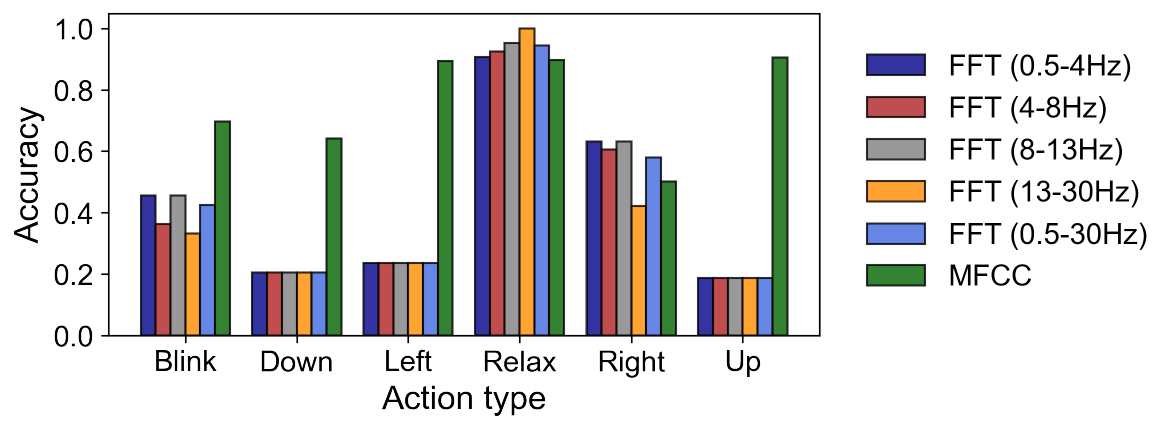}\\
  \caption{The Comparison in artifact recognition accuracy using MFCC or FFT.}
  \label{fig:comparison-FFT}
\end{figure}

In order to evaluate the effectiveness of the cepstral coefficients, a comparison between the Fast Fourier Transform (FFT) and the cepstral analysis based features was carried out.
The FFT is used for feature extraction on the collected EEG signals and the training of several classifiers on five frequency bands (0.5-4Hz, 4-8Hz, 8-13Hz, 13-30Hz, 0.5-30Hz), respectively.
However, in the same case of using 5-fold CV and grid search, the accuracy of the artifact identification using FFT as the feature extraction method is much lower than that of using the proposed method in 4 of 6 actions as shown in Fig. \ref{fig:comparison-FFT}.
The result indicates that when more calculations than FFT is paid, better performance of artifact recognition is achieved by using cepstral analysis.

\subsection{Artifact Removal}
According to the deconvolution characteristic of cepstral analysis, the clean EEG and artifacts can be separated in cepstral domain, which means some dimensions of the MFCCs may be dominated by artifacts, while other dimensions represent the clean EEG signals.
The Pearson correlation coefficients of the samples with artifacts and the clean samples in each dimension can be calculated as
\begin{small}
\begin{equation}
\label{eq:pearson}
\begin{aligned}
Cor(A^{(i)},R^{(i)})=\frac{\displaystyle\sum_{j=1}^N({A_j}^{(i)}-\Bar{A}^{(i)})({R_j}^{(i)}-\Bar{R}^{(i)})}{\displaystyle\sqrt{{\sum_{j=1}^N{({A_j}^{(i)}-\Bar{A}^{(i)})}^2}{\sum_{j=1}^N{({R_j}^{(i)}-\Bar{R}^{(i)})}^2}}}
\end{aligned}
\end{equation}
\end{small}

\noindent {where $A^{(i)}$ and $R^{(i)}$ are the $i$-th dimensions of the artifact samples and the clean samples. $\Bar{A}^{(i)}$ and $\Bar{R}^{(i)}$ represent the mean of the artifact samples and clean samples of the $i$-th dimensions, respectively.
$N$ is the number of clean samples.}

Considering the compromise of the removal effect and the information loss, dimensions with the smallest two coefficients (the 11th- and the 12th-dimension in this work) in the artifact samples are considered to be dominated by the artifacts. These two coefficients are transformed as 

\begin{equation}
\label{eq:removal}
\begin{aligned}
\hat{A}^{(i)}=\frac{A^{(i)}-\Bar{A}^{(i)}}{\sigma_{A^{(i)}}}\times{\sigma_{R^{(i)}}}+\Bar{R}^{(i)}
\end{aligned}
\end{equation}
where $\sigma$ is the standard deviation. 
The trained binary SVM classifier is used to classify the transformed samples. The result shows an artifact rate of 9$\%$, which means that 91$\%$ of the artifact samples were successfully denoised.

\begin{figure}[htb]
  \centering
  \includegraphics[width=0.48\textwidth]{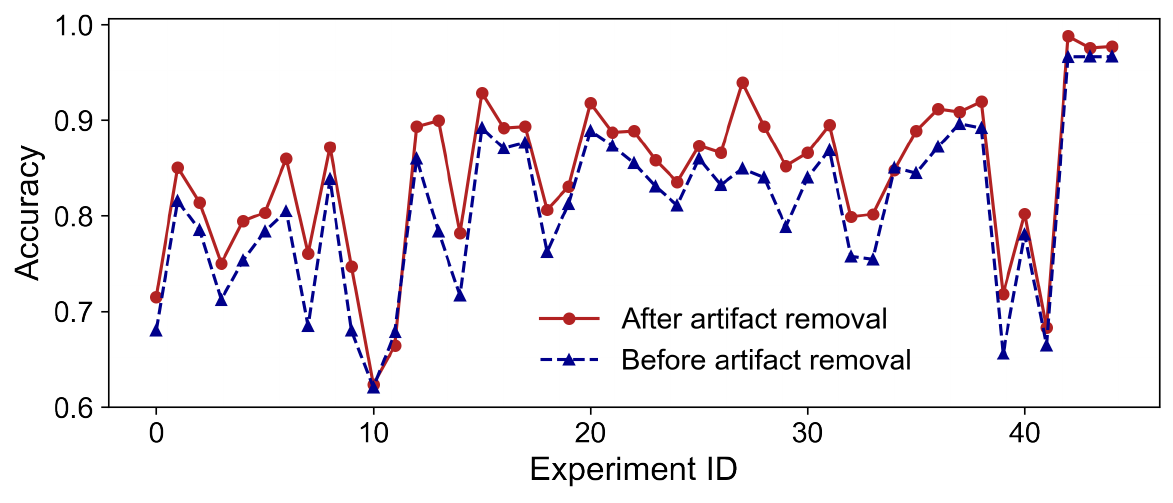}\\
  \caption{Evaluation of the artifact removal using MFCCs. An average accuracy increasing of 3.46\% is monitored.}
  \label{fig:seed}
\end{figure}

The proposed artifact removal method using the cepstral features was evaluated on SJTU Emotion EEG Dataset (SEED) \cite{seed-2015}, a public EEG database for emotion recognition.
There are 15 subjects with 45 experiments. In the evaluation, the differential entropy (DE) \cite{duan2013de} features were utilized together with a SVM classifier. The 22 EEG channels acquired closed to the forehead were used, which are the channel indexes from 0 to 21 in SEED database. The EEG signals in SEED were exploited to get the baseline emotion recognition performance, where an overall average 3-category classification accuracy of 80.92\% was achieved.
The proposed cepstral feature based artifact removal method was applied to the denoise of the target EEG signals.
Since there is no label information of the artifacts, the 11th- and the 12th-dimension of the MFCCs were removed directly according to the analysis results in our labeled dataset. The processed MFCCs were used to reconstruct the denoised EEG. With the denoised EEG, an overall average accuracy of 84.38\% was achieved. According to Fig. \ref{fig:seed}, the reconstructed EEG signals perform better in 43 of 45 experiments, where an average accuracy increase of 3.46\% is obtained.
The whole process of the artifact removal method was described in Algorithm \ref{alg:removal}.

\begin{algorithm}[htb]
  \small
  \caption{The process of artifact removal method.}
 \label{alg:removal}
 \begin{algorithmic}[1]
  \Require
    \\
     Source EEG database with labeled artifact, $EEG_{source}$ (i.e., the recorded EEG in this work);
     \\
     Target EEG database, $EEG_{target}$ (e.g., SEED database);
  \Ensure
  \\
     The denoised EEG signals of the target database, $EEG_{target-denoised}$; 
  \renewcommand{\algorithmicrequire}{\textbf{Find the artifact         related dimensions in cepstral domain:}}
  \Require
   \For{all clean EEG segment $EEG_{source-clean}^{i}$ in source database}
   \State Calculate the MFCCs $MFCC_{source-clean}^{i}$.
   \EndFor
   \For{all artifacted EEG segment $EEG_{source-artifact}^{i}$ in source database}
   \State Calculate the MFCCs $MFCC_{source-artifact}^{i}$.
   \EndFor
   \State Use Equation \ref{eq:pearson} to get the Pearson correlation coefficients $Cor_{1:12}$ between artifacted samples and clean samples.
   \State Sort the $Cor_{1:12}$ and pick the smallest two as artifact related dimensions $Dim_{artifact}^{1}$ and $Dim_{artifact}^{2}$.
   \renewcommand{\algorithmicrequire}{\textbf{Artifact removal in target database:}}
   \Require
   \For{all EEG segment $EEG_{target}^{i}$ in target database}
   \State Calculate the MFCCs $MFCC_{target}^{i}$.
   \State{\textbf{if} There are artifact labels and clean labels in target database \textbf{then}}
   \State {\quad Use Equation \ref{eq:removal} to normalize the artifact related} 
   \State {\quad dimensions $Dim_{artifact}^{1}$ and $Dim_{artifact}^{2}$}
   \State{\textbf{else} \textbf{if} There is only EEG data in target database \textbf{then}}
   \State {\quad Remove the two artifact related dimensions $Dim_{artifact}^{1}$} 
   \State {\quad and $Dim_{artifact}^{2}$.}
   \State{\textbf{end if}}
   \State Reconstruct the corresponding EEG segments $EEG_{target-denoised}^{i}$ using $MFCC_{target}^{i}$.
   \EndFor \\
   \Return $EEG_{target-denoised}$
 \end{algorithmic}
\end{algorithm}

\subsection{Comparison with the state-of-the-art}
The proposed method is compared with several state-of-the-art solutions as shown in Table \ref{tb:state-of-the-art}.
The computational cost is estimated according to the number of multiplication operations in the unit of frame.
\begin{table}[!ht]
\caption{Comparison with several state-of-the-art methods}
\label{tb:state-of-the-art}
\begin{center}
\scriptsize
\begin{threeparttable}
\scalebox{0.95}{
\begin{tabular}{|c|c|c|c|}
\hline
\textbf{Methods} & \textbf{Accuracy} & \textbf{Computational cost} &  \textbf{Removal} \\
\hline
ICA-DWT (2 thresholds) \cite{mahajan-ica-JBHI2014} & 39.00\% & 6.32$\times{10^{9}}$ & Yes\\
\hline
DWT-ICA (multi-thresholds) \cite{daly2014force} & 64.33\% & 7.89$\times{10^{9}}$ & Yes\\
\hline
CNN \cite{khatwani-biocas18} & 73.60\%\tnote{a} & 1.26$\times{10^{7}}$ & No \\
\hline
LSTM \cite{manjunath2020low} & 93.10\%\tnote{a} & 4.15$\times{10^{5}}$ & No \\
\hline
\textbf{Our Method} & \textbf{99.62\%} & \textbf{6.56\bm{$\times{10^{5}}$}} & \textbf{Yes} \\
\hline
\end{tabular}}
 \begin{tablenotes}
        \footnotesize
        \item[a] Database shift
      \end{tablenotes}
    \end{threeparttable}
\end{center}
\end{table}
In order to calculate the artifact detection accuracy, two typical automatic hybrid algorithms \cite{mahajan-ica-JBHI2014, daly2014force} combining ICA with DWT were re-implemented and evaluated using the recorded EEG in this work.
The original accuracies of \cite{khatwani-biocas18} and \cite{manjunath2020low} were reported directly since channel number mismatch.
Results show the proposed method features higher accuracy and lower computational cost than the traditional BSS based methods \cite{mahajan-ica-JBHI2014, daly2014force} in same dataset.
Although widely recognized in off-line artifact removal, since the lack of confident statistic values, the BSS based methods are difficult to handle a short-term EEG segment, which makes them unsuitable to be integrated in real-time BMIs.
In the field of supervised methods, the proposed method as well as \cite{khatwani-biocas18} and \cite{manjunath2020low} provides a reasonable high detection accuracy.

The computational complexity of the two fundamental BSS modules, ICA and CCA, is estimated to be $O(3pC^{2}N)$ and $O(CN^{3})$, where $p$ is the iterations number of FastICA. $C$ is the number of signal channels. $N$ is the number of samples in each frame.
In the proposed method, the algorithm complexity mainly comes from the calculation of the FFT, the mel filter and the DCT.
The FFT reduces the complexity of the calculation of the DFT from $O(N^2)$ to $O(NlogN)$.
For the calculation of the output of each mel filter, a total of $M\times{N}$ multiplications are required, where $M$ is the number of the mel filters. In addition, the calculation of the result of the DCT requires $M\times{L}$ multiplications, where $L$ is the coefficient order of MFCCs. 
The above calculation are required by each channel. The total multiplication times is proportional to the number of the channels $C$. As a result, the proposed artifact detection method features a time complexity of $O(C(\frac{N}{2}log_{2}N+MN))$, which is usually smaller than that of the ICA and the CCA at the frame level ($N<p$).
In Table \ref{tb:state-of-the-art}, a typical iteration number of 20k was used to estimate the computational cost for ICA.
According to the comparison, the proposed method features a significant lower computational cost than most existed solutions in literature, especially the BSS family.
The neural network (NN) family also features low computational cost since neither iteration nor spectrum transform is required. 
However, the NN based methods have difficulties in artifact removal.
Compared to the FBSE-EWT \cite{bhattacharyya2018fourier} based method for ocular artifact removal \cite{gajbhiye2019elimination}, the proposed method holds a filtering free characteristic and offers the abilities of artifact detection and recognition.

\section{Conclusion}
This paper introduced the cepstrum into the fields of ocular artifact detection, recognition and removal for prefrontal EEG.
Based on cepstral feature, a low computational cost, high accuracy, reference free automatic artifact detection and recognition method was developed.
The proposed method achieved a 99.62\% accuracy in the artifact detection task and an 82.79\% accuracy in the 6-class artifact recognition task.
In addition, a statistic value-based artifact removal algorithm was proposed for the scenarios where the contaminated EEG should not be dropped out. Evaluation results show that the proposed artifact removal method improved the classification accuracy of 3.46\% on a public emotion recognition EEG database. 
The proposed solution features a high artifact detection and recognition accuracy as well as a low computational cost compared to the existing solutions, which is suitable for real-time BMI applications.




\footnotesize 
\begin{thebibliography}{99}

\bibitem{tang-nature}
M.~Zhang et al., ``Electronic neural
  interfaces,'' \emph{Nature Electronics}, vol.~3, no.~4, pp. 191--200, 2020.

\bibitem{rssi-seizure}
Y.~Zhang et al., ``An ultra-low power
  RSSI amplifier for EEG feature extraction to detect seizures,'' \emph{IEEE Transactions 
  on Circuits and Systems II: Express Briefs}, vol.~69, no.~2, pp. 329--333, 2021.

\bibitem{mcmd-sleep}
C.~Zhang et al., ``Multichannel Multidomain-Based Knowledge Distillation 
  Algorithm for Sleep Staging With Single-Channel EEG,'' \emph{IEEE Transactions 
  on Circuits and Systems II: Express Briefs}, vol.~69, no.~11, pp. 4608--4612, 2022.


\bibitem{jiang2019removal-review}
X.~Jiang et al., ``Removal of artifacts from EEG signals: a
  review,'' \emph{Sensors}, vol.~19, no.~5, p. 987, 2019.

\bibitem{zou2019removing}
L.~Zou et al., ``Removing muscle artifacts from EEG data via underdetermined joint blind source separation: A simulation study,'' \emph{IEEE Transactions on Circuits and Systems II: Express Briefs}, vol.~67, no.~1, pp. 187--191, 2019.

\bibitem{mahajan-ica-JBHI2014}
R.~Mahajan et al., ``Unsupervised eye blink artifact denoising of
  EEG data with modified multiscale sample entropy, kurtosis, and
  wavelet-ICA,'' \emph{IEEE journal of Biomedical and Health Informatics},
  vol.~19, no.~1, pp. 158--165, 2014.

\bibitem{daly2014force}
I.~Daly et al., ``Force: Fully
  online and automated artifact removal for brain-computer interfacing,''
  \emph{IEEE transactions on neural systems and rehabilitation engineering},
  vol.~23, no.~5, pp. 725--736, 2014.

\bibitem{lee2018removal}
S.~Lee et al., ``Removal of high-voltage brain
  stimulation artifacts from simultaneous EEG recordings,'' \emph{IEEE
  Transactions on Biomedical Engineering}, vol.~66, no.~1, pp. 50--60, 2018.

\bibitem{cimmino-pca-2021}
A.~Cimmino et al., ``Non-linear PCA neural
  network for EEG noise reduction in brain-computer interface,'' in
  \emph{Progresses in Artificial Intelligence and Neural Systems}.\hskip 1em
  plus 0.5em minus 0.4em\relax Springer, 2021, pp. 405--413.

\bibitem{asr-tcasii}
L.~Van et al., ``Hardware-Oriented Memory-Limited Online Artifact 
  Subspace Reconstruction (HMO-ASR) Algorithm,'' \emph{IEEE Transactions 
  on Circuits and Systems II: Express Briefs}, vol.~68, no.~12, pp. 3493--3497, 2021.

\bibitem{chen-cca-2018}
X.~Chen et al., ``A novel EEMD-CCA approach to
  removing muscle artifacts for pervasive EEG,'' \emph{IEEE Sensors Journal},
  vol.~19, no.~19, pp. 8420--8431, 2018.

\bibitem{2019-filter}
C.~Dai et al., ``Removal of ECG artifacts
  from EEG using an effective recursive least square notch filter,'' \emph{IEEE
  Access}, vol.~7, pp. 158\,872--158\,880, 2019.

\bibitem{khatun2016-wavelet}
S.~Khatun et al., ``Comparative study of wavelet-based
  unsupervised ocular artifact removal techniques for single-channel EEG
  data,'' \emph{IEEE journal of translational engineering in health and
  medicine}, vol.~4, pp. 1--8, 2016.

\bibitem{2020-JBHI-wavelet}
S.~Phadikar et al., ``Automatic eyeblink artifact removal from
  EEG signal using wavelet transform with heuristically optimized threshold,''
  \emph{IEEE Journal of Biomedical and Health Informatics}, vol.~25, no.~2, pp.
  475--484, 2020.

\bibitem{gajbhiye2019elimination}
P.~Gajbhiye et al., ``Elimination of ocular
  artifacts from single channel eeg signals using fbse-ewt based rhythms,''
  \emph{IEEE Sensors Journal}, vol.~20, no.~7, pp. 3687--3696, 2019.

\bibitem{khatwani-biocas18}
M.~Khatwani et al., ``Energy efficient convolutional neural networks for EEG
  artifact detection,'' in \emph{2018 IEEE Biomedical Circuits and Systems
  Conference (BioCAS)}.\hskip 1em plus 0.5em minus 0.4em\relax IEEE, 2018, pp.
  1--4.

\bibitem{manjunath2020low}
N.~K. Manjunath et al., ``A low-power LSTM processor for multi-channel brain EEG
  artifact detection,'' in \emph{2020 21st International Symposium on Quality
  Electronic Design (ISQED)}.\hskip 1em plus 0.5em minus 0.4em\relax IEEE,
  2020, pp. 105--110.

\bibitem{deep-learning2020}
J.~R. McIntosh et al., ``Ballistocardiogram
  artifact reduction in simultaneous EEG-fMRI using deep learning,'' \emph{IEEE
  Transactions on Biomedical Engineering}, vol.~68, no.~1, pp. 78--89, 2020.

\bibitem{childers1977cepstrum}
D.~G. Childers et al., ``The cepstrum: A guide to
  processing,'' \emph{Proceedings of the IEEE}, vol.~65, no.~10, pp.
  1428--1443, 1977.

\bibitem{juvela2018-ICASSP18}
L.~Juvela et al., ``Speech waveform synthesis from MFCC sequences with generative
  adversarial networks,'' \emph{ICASSP}.\hskip 1em plus 0.5em minus
  0.4em\relax IEEE, 2018, pp. 5679--5683.

\bibitem{luo2019low}
D.~Luo et al., ``A low-noise chopper amplifier designed for
  multi-channel neural signal acquisition,'' \emph{IEEE Journal of Solid-State
  Circuits}, vol.~54, no.~8, pp. 2255--2265, 2019.

\bibitem{mfcc1980}
S.~Davis et al., ``Comparison of parametric representations for monosyllabic word recognition in continuously spoken sentences,'' \emph{IEEE transactions on acoustics, speech, and signal processing}, vol.~28, no.~4, pp. 357--366, 1980.

\bibitem{seed-2015}
W.-L. Zheng et al., ``Investigating critical frequency bands and channels
  for EEG-based emotion recognition with deep neural networks,'' \emph{IEEE
  Transactions on Autonomous Mental Development}, vol.~7, no.~3, pp. 162--175,
  2015.

\bibitem{duan2013de}
R.-N. Duan et al., ``Differential entropy feature for
  EEG-based emotion classification,'' \emph{IEEE/EMBS
  Conference on Neural Engineering (NER)}.\hskip 1em plus 0.5em minus
  0.4em\relax IEEE, 2013, pp. 81--84.

\bibitem{bhattacharyya2018fourier}
A.~Bhattacharyya et al., ``Fourier--bessel series
  expansion based empirical wavelet transform for analysis of non-stationary
  signals,'' \emph{Digital Signal Processing}, vol.~78, pp. 185--196, 2018.

\end{thebibliography}

\end{document}